\documentclass[epj]{svjour}
\usepackage{graphicx}

\def\mb#1{\setbox0=\hbox{$#1$}\kern-.025em\copy0\kern-\wd0
\kern-0.05em\copy0\kern-\wd0\kern-.025em\raise.0233em\box0}

\begin{document}
   \title{Curious behaviour of the diffusion coefficient and friction force for the strongly
   inhomogeneous HMF model}

 \author{P.H. Chavanis}

\institute{ Laboratoire de Physique Th\'eorique, Universit\'e Paul
Sabatier, 118 route de Narbonne 31062 Toulouse, France\\
\email{chavanis@irsamc.ups-tlse.fr} }

\titlerunning{Strongly inhomogeneous HMF model}

   \date{To be included later }

   \abstract{We present first elements of kinetic theory appropriate
   to the inhomogeneous phase of the Hamiltonian Mean Field (HMF)
   model. In particular, we investigate the case of strongly
   inhomogeneous distributions for $T\rightarrow 0$ and exhibit
   curious behaviour of the force auto-correlation function and
   friction coefficient. The temporal correlation function of the
   force has an oscillatory behaviour which averages to zero over a
   period. By contrast, the effects of friction accumulate with time
   and the friction coefficient does not satisfy the Einstein
   relation. On the contrary, it presents the peculiarity to increase
   linearly with time. Motivated by this result, we provide analytical
   solutions of a simplified kinetic equation with a time dependent
   friction. Analogies with self-gravitating systems and other systems
   with long-range interactions are also mentioned.
 }

   \maketitle
%

\section{Introduction}
\label{sec_introduction}

Recently, there was a renewed interest for the statistical mechanics
of systems with long-range interactions \cite{dauxois}. This concerns
self-gravitating systems (galaxies) in astrophysics
\cite{paddy,houches}, large-scale coherent structures (jets and
vortices) in geophysical flows \cite{houches}, bacterial populations
(chemotaxis) in biology \cite{murray,ribot}, clusters in the
Hamiltonian Mean Field (HMF) and Brownian Mean Field (BMF) models
\cite{hmf,vatt}, galactic bars \cite{pichon}, neutral and non-neutral plasmas,
dislocation dynamics, planetary formation, cosmology etc. The
dynamical evolution of such systems presents a lot of peculiarities
\cite{bh}. For Hamiltonian systems with long-range interactions, the
collisional relaxation time diverges with the number $N$ of particles
so that the system experiences two successive types of relaxation: a
{\it collisionless relaxation} on a short timescale of the order of a
few dynamical times $t_{D}$ (called violent relaxation in
astrophysics) and a {\it collisional relaxation} on a long time scale
of the order $N^{\delta}t_{D}$ with $\delta\ge 1$. The first regime
leads to the formation of a metaequilibrium state, or 
quasi-stationary state (QSS), which is a stable stationary solution of
the Vlasov equation that is not necessarily of the Boltzmann form
\cite{lb,tremaine,csr,yamaguchi,cstsallis,super,next5}. 
The second regime leads in general to the ordinary statistical
equilibrium state described by the Boltzmann distribution. In the case
of self-gravitating systems, there may not exist statistical
equilibrium and the system can evaporate or collapse (gravothermal
catastrophe) \cite{lbw}. Between the phase of violent relaxation and the late
collisional evolution (equilibrium or collapse) the system passes by a
succession of quasi-stationary states which are quasi-stationary
solutions of the Vlasov equation slowly evolving with time under the
effect of encounters (finite $N$ effects). In
astrophysics, this phase is described by the
orbit-averaged-Fokker-Planck equation \cite{bt,cstsallis}.

The developement of a kinetic theory for the collisional evolution of
systems with long-range interactions is complicated for different
reasons. First of all, the temporal correlation function of the force
may not decay sufficiently rapidly to vindicate the Markovian
approximation that is used in many kinetic theories. For example, in
stellar dynamics, the temporal correlation function decreases like
$t^{-1}$ leading to a logarithmic divergence of the diffusion
coefficient \cite{chandra}. In the case of the HMF model, the
correlation function decreases exponentially rapidly but the
correlation time diverges close to the critical point $T\rightarrow
T_{c}$ \cite{bouchet,vatt}. This is a general feature of long-range
attractive potentials of interaction \cite{bh}. On the other hand,
systems with long-range interactions are usually spatially {\it
inhomogeneous} and non-local effects strongly complicate the kinetic
theory. In the case of self-gravitating systems, one usually avoids
the problem by making a local approximation and developing the kinetic
theory as if the system were homogeneous \cite{bt}. This is partly
justified by the fact that the fluctuations of the gravitational force
are dominated by the contribution of the nearest neighbor (the
distribution of the force is a particular L\'evy law called the
Holtzmark distribution)
\cite{channeu}. On the other hand, for the HMF model, the kinetic
theory has been developed only in the case $T>T_{c}$ where the system
is in a stable homogeneous phase
\cite{bd,vatt,cl}. In these situations the meanfield force vanishes and,
to a first approximation, the particles follow linear trajectories
with constant velocity.

One goal of this paper is to present elements of kinetic theories
valid for the inhomogeneous phase of the HMF model ($T<T_{c}$) and
show that the situation becomes sensibly different from what we are
used to in the case of homogeneous systems. In particular, we shall
investigate the case of strongly inhomogeneous systems for
$T\rightarrow 0$, where the particles cluster around $\theta=0$. In
that case, their mean motion is that of a harmonic oscillator and it
is possible to calculate analytically the auto-correlation function of
the force and the friction. We find that these quantities present a
curious behaviour. The temporal correlation function of the force has
an oscillatory evolution which averages to zero over a period. By
contrast, the effects of friction accumulate with time and the
friction coefficient does not satisfy the Einstein relation. On the
contrary, it presents the peculiarity to increase linearly with time.
These curious behaviours were previously noted by Kandrup
\cite{kandrup2} in the case of self-gravitating systems but the consideration
of the HMF model allows to obtain more explicit results (devoid of any
gravitational divergences) and provides a simple framework where these
effects can be studied in detail.

The paper is organized as follows. In Sec. \ref{sec_stat}, we recall
basic results concerning the structure of the statistical equilibrium
states of the HMF model. In Sec. \ref{sec_in}, we develop a kinetic
theory of the HMF model valid for both the homogeneous and the
inhomogeneous phase. We present a generalized Landau equation
describing the evolution of the distribution function of the system as
a whole and a generalized Fokker-Planck equation describing the
evolution of the distribution function of a test particle (or an
ensemble of non-interacting test particles) in a thermal bath of field
particles at statistical equilibrium. In Sec. \ref{sec_auto}, we
calculate the temporal auto-correlation function of the force acting
on the test particle. We show that it presents an oscillatory behavior
which averages to zero over a period. We also discuss the expression
of the diffusion coefficient and its relation to the Kubo formula. In
Sec. \ref{sec_fric}, we calculate the friction force acting on the
test particle. We show that the frictional effects are cumulative and
lead to a linear divergence of the friction coefficient for
$t\rightarrow +\infty$. We also discuss the break-up of the Einstein
relation for a strongly inhomogeneous system. In Sec. \ref{sec_time},
we provide the analytical solution of a simplified kinetic equation
with a time-dependent friction related to our study.

\section{Statistical equilibrium states of the HMF model}
\label{sec_stat}

The HMF model consists of $N$ particles (of unit mass) moving on a ring and
interacting via a cosinusoidal potential. The phase space coordinates
$(\theta_{i},v_{i})$ of the particles satisfy the Hamiltonian equations
of motion
\begin{eqnarray}
\label{stat1}
{d\theta_{i}\over dt}={\partial H\over\partial v_{i}}, \qquad {dv_{i}\over dt}=-{\partial H\over\partial \theta_{i}},\nonumber\\
H=\sum_{i=1}^{N}{1\over 2}v_{i}^{2}-{k\over 4\pi}\sum_{i\neq j}\cos(\theta_{i}-\theta_{j}).
\end{eqnarray}
The HMF model was introduced by several groups (see a short historic in
\cite{vatt}) either as a simple model with long-range interactions
mimicking gravitational dynamics \cite{konishi,inagaki,ruffo} or as a
simplified model for the formation of bars in disk-shape galaxies
\cite{pichon}. The thermodynamic limit corresponds to $N\rightarrow +\infty$
in such a way that the rescaled temperature $\eta=kM/4\pi T$ and
rescaled energy $\epsilon=8\pi E/kM^{2}$ remain of order unity (this
can be conveniently accomplished by taking the coupling constant
$k\sim 1/N$ right from the begining, in which case $T\sim 1$ and
$E\sim N$). In this proper thermodynamic limit, the mean-field
approximation becomes exact for $N\rightarrow +\infty$, except near
the critical point \cite{bh}. At statistical equilibrium (see, e.g.,
\cite{vatt}), the distribution function can be written
$f(\theta,v)=\rho(\theta)f(v)$ with
\begin{eqnarray}
\label{stat2}
f(v)=\left (\frac{\beta}{2\pi}\right )^{1/2}e^{-\beta \frac{v^{2}}{2}},
\end{eqnarray}
\begin{eqnarray}
\label{stat3}
\rho(\theta)=\frac{M}{2\pi I_{0}(\beta B)}e^{-\beta B\cos\theta},
\end{eqnarray}
where we have adopted the normalization
$\int_{0}^{2\pi}\rho(\theta)d\theta=M$ and
$\int_{-\infty}^{+\infty}f(v)dv=1$ (note that here $M=Nm=N$). Equation
(\ref{stat3}) is the Boltzmann distribution in a potential
$\Phi=B\cos\theta$. The meanfield force experienced by a particle is
\begin{eqnarray}
\label{stat4}
\langle F\rangle (\theta)=-\Phi'(\theta)=B\sin\theta, 
\end{eqnarray}
where
\begin{eqnarray}
\label{stat5}
B=-\frac{k}{2\pi}\int_{0}^{2\pi}\rho(\theta')\cos\theta' d\theta'.
\end{eqnarray}
To obtain Eqs. (\ref{stat4})-(\ref{stat5}), we have assumed (without
loss of generality) that the equilibrium distribution of the system is
symmetric with respect to the $x$-axis. The quantity $B$ is similar to
the magnetization in spin systems.  It is determined as a function of
the temperature (see e.g. \cite{vatt}) by the implicit equation
\begin{eqnarray}
\label{stat6}
B=\frac{kM}{2\pi}\frac{I_{1}(\beta B)}{I_{0}(\beta B)}.
\end{eqnarray}
The energy is given by $E=NT/2-\pi B^{2}/k$. For $T>T_{c}=kM/4\pi$ or
$E>E_{c}=kM^{2}/8\pi$, the only solution to the above equation is
$B=0$ leading to an homogeneous distribution of particles. For
$T<T_{c}$ or $E<E_{c}$, the homogeneous phase becomes unstable and a
(stable) clustered phase appears with $B\neq 0$. This corresponds to a
second order phase transition (see e.g. \cite{vatt}).

\section{The inhomogeneous kinetic  equation}
\label{sec_in}

We shall discuss here the kinetic theory of the HMF model by using
general results coming from the projection operator formalism. This
formalism starts from the Liouville equation for the $N$-body
distribution function $P_{N}(\lbrace \theta_{i}, v_{i}\rbrace ,t)$ and
derives an exact kinetic equation for the one-body distribution
function $f(\theta,v,t)=NP_{1}(\theta,v,t)$ by using projection
technics. This equation is then simplified by making some
approximations on the correlation function of the field particles.
This formalism introduced by Willis \& Picard \cite{wp} is quite
general and  leads to a form of generalized Landau equation
\cite{bh}. It was applied by Kandrup \cite{kandrup1} in the case of 
stellar systems, by Chavanis \cite{kinvortex,houches} for
two-dimensional point vortices and in \cite{vatt} for the HMF
model. This formalism is also very close to the linear response theory
developed in \cite{kandrup2,rapid} where the friction term (or the drift
term in the case of point vortices) is calculated directly from the
perturbation of the $N$-body distribution function of the bath caused
by the interaction with the test particle. One interest of this
formalism is that it remains valid in the case of non-Markovian and
spatially inhomogeneous systems while other formalisms developed in
connection with plasma physics (binary collision models, BBGKY
hierarchy, diagrammatic expansions, quasilinear theory,...) usually
consider Markovian and homogeneous systems and work in Fourier
space. By contrast, the projection operator formalism remains in
physical space which enlightens the basic physics. We shall not repeat
the intermediate steps of the formalism which can be found in
\cite{wp,kandrup1,kinvortex}. This
formalism leads to a general kinetic equation of the form 
\begin{eqnarray}
\label{in1}
\frac{\partial f}{\partial t}+v\frac{\partial f}{\partial \theta}+\langle F\rangle \frac{\partial f}{\partial v}=\frac{\partial}{\partial v}\int_{0}^{t}d\tau\int d\theta_{1}dv_{1} {\cal F}(1\rightarrow 0,t)\nonumber\\
\times G(t,t-\tau) \left \lbrack {\cal F}(1\rightarrow 0,t-\tau) \frac{\partial }{\partial v}+{\cal F}(0\rightarrow 1,t-\tau)\frac{\partial}{\partial v_{1}}\right \rbrack\nonumber\\
\times f(\theta,v,t-\tau)f(\theta_{1},v_{1},t-\tau), \qquad
\end{eqnarray}
where $G(t,t-\tau)$ is the Green function associated to the averaged
Liouville operator constructed with the meanfield force $\langle
F\rangle (\theta,t)$ and ${\cal F}(1\rightarrow 0,t)$ is the
fluctuating force created by particle $1$ on particle $0$ (see, e.g.,
\cite{kandrup1} for more details). Under this form, we clearly see 
the terms of diffusion and friction (first and second terms in the
r.h.s. of Eq. (\ref{in1})) and their connection to a generalized form
of Kubo formula (the time integral of the force auto-correlation
function). These points will be developed in the sequel. The ratio of
the right hand side (collision term) on the left hand side (meanfield
advective term) is of order $1/N$ in the proper thermodynamic limit
\cite{vatt,bh} recalled in Sec. \ref{sec_stat}. Therefore, for $N\rightarrow
+\infty$, Eq. (\ref{in1}) reduces to the Vlasov equation. The
collision term takes into account finite $N$ effects and can be viewed
as the first order correction to the Vlasov limit in an expansion in
$N^{-1}$ (see, e.g.,
\cite{houches} p. 260).  Equation (\ref{in1}) can be viewed as a
generalization of the Landau equation (initially introduced in plasma
physics) to which it reduces \footnote{Note, however, that for
one-dimensional systems such as the HMF model, the Landau collision
term cancels out \cite{bd,vatt,bh}; see also last paragraph of \cite{kp}.} if the system is homogeneous and
Markovian \cite{bh}. One drawback, however, of the projection operator
formalism (or more precisely of the approximations leading to
Eq. (\ref{in1})) is that it ignores collective effects which are
important especially close to the critical point. Such collective
effects can be taken into account by using the Lenard-Balescu equation
in the case of homogeneous systems
\cite{bd,vatt,bh}. We shall not discuss these collective
effects here  and shall remain close to the situation considered by
Kandrup \cite{kandrup1} in the astrophysical setting by adapting and
expliciting the calculations in the case of the HMF model.

Equation (\ref{in1}) is an integrodifferential equation (with respect
to the variables $\theta_{1}$, $v_{1}$) describing the evolution of
the system as a whole.  We shall consider here a simpler problem,
namely the evolution of a test particle in a bath of field particles
with {\it prescribed} static distribution $f(\theta_{1},v_{1})$ which
is a stable stationary solution of the Vlasov equation. By adapting
the projection operator formalism to this situation where
$f(\theta_{1},v_{1})$ is fixed, we find that the time evolution of the
density probability $P(\theta,v,t)$ of finding the test particle in
$(\theta,v)$ at time $t$ is governed by the equation
\begin{eqnarray}
\label{in2}
\frac{\partial P}{\partial t}+v\frac{\partial P}{\partial \theta}+\langle F\rangle \frac{\partial P}{\partial v}=\frac{\partial}{\partial v}\int_{0}^{t}d\tau\int d\theta_{1}dv_{1} {\cal F}(1\rightarrow 0,t)\nonumber\\
\times \left \lbrack {\cal F}(1\rightarrow 0,t-\tau)  \frac{\partial }{\partial v}+{\cal F}(0\rightarrow 1,t-\tau)  \frac{\partial}{\partial v_{1}}\right \rbrack\nonumber\\
\times P(\theta(t-\tau),v(t-\tau),t-\tau) f(\theta_{1}(t-\tau),v_{1}(t-\tau)). \qquad 
\end{eqnarray}
 In this paper, we shall consider the evolution of a test particle in a {\it thermal bath} of field particles at statistical equilibrium described by the distribution (\ref{stat2})-(\ref{stat3}). In that case, we obtain a kinetic equation of the form
\begin{eqnarray}
\label{in4}
\frac{\partial P}{\partial t}+v\frac{\partial P}{\partial \theta}+\langle F\rangle \frac{\partial P}{\partial v}=\frac{\partial}{\partial v}\int_{0}^{t}d\tau\int d\theta_{1}dv_{1} {\cal F}(1\rightarrow 0,t)\nonumber\\
\times \biggl\lbrace  {\cal F}(1\rightarrow 0,t-\tau)\frac{\partial }{\partial v}
-{\cal F}(0\rightarrow 1,t-\tau)\beta v_{1}(t-\tau)\biggr\rbrace \nonumber\\
\times P(\theta(t-\tau),v(t-\tau),t-\tau)\rho(\theta_{1})f(v_{1}).\qquad
\end{eqnarray}
The fluctuating force can be written ${\cal F}(1\rightarrow 0,t)=F(1\rightarrow 0,t)-\langle F(1\rightarrow 0,t) \rangle$ where
\begin{eqnarray}
\label{in3}
F(1\rightarrow 0,t)=-\frac{k}{2\pi}\sin(\theta(t)-\theta_{1}(t)),
\end{eqnarray}
is the exact value of the force created by particle $1$ on particle $0$ 
and
\begin{eqnarray}
\label{inadd1}
\langle {F}(1\rightarrow 0,t)\rangle=\frac{1}{N}\int {F}(1\rightarrow 0,t) \rho(\theta_{1})f(v_{1})d\theta_{1}dv_{1},\nonumber\\ 
\end{eqnarray}
is its mean-field value.  Equation (\ref{in4}) can be seen as a sort
of generalized Fokker-Planck equation. However, the dynamics is
generally non-Markovian (see below) so that Eq. (\ref{in4}) is {\it
not} a Fokker-Planck equation.

The time integral in Eq. (\ref{in4}) must be performed by moving the
particles with the meanfield force (\ref{stat4}) between $t$ and
$t-\tau$ (see, e.g., \cite{kandrup1}). Accordingly, the quantities
$\theta(t-\tau)$ and $v(t-\tau)$ are solutions of the equation of
motion
\begin{eqnarray}
\label{in5}
\frac{d^{2}\theta}{dt^{2}}=B\sin\theta.
\end{eqnarray}
This is the equation of a nonlinear oscillator. The general solution 
is given by
\begin{eqnarray}
\label{in6}
\int_{\theta_{0}}^{\theta(t)}\frac{d\phi}{\sqrt{v_{0}^{2}+2B(\cos\theta_{0}-\cos\phi)}}=\pm t,
\end{eqnarray}
\begin{eqnarray}
\label{in7}
v(t)=\pm \sqrt{v_{0}^{2}+2B(\cos\theta_{0}-\cos\theta(t))}.
\end{eqnarray}
In our previous study \cite{vatt}, we have considered the case
$T>T_{c}$ where the distribution of the bath is homogeneous and the
particles follow linear trajectory with constant velocity in a first
approximation. In that case, Eq. (\ref{in4}) reduces to a
Fokker-Planck equation of the Kramers form with a friction coefficient
given by the Einstein relation.  This kinetic equation converges for
$t\rightarrow +\infty$ to the Maxwellian distribution of the bath
\cite{vatt}. However, the diffusion coefficient decreases rapidly with
the velocity (like the Gaussian distribution of the bath) and this
leads to anomalous diffusion
\cite{bd} and to a slow progression of the front in the high velocity
tail \cite{cl}. Here, we shall consider the case $T\rightarrow 0$
where the bath distribution is strongly inhomogeneous. Assuming
$B=-\omega^{2}<0$ (the case $B>0$ is symmetric), the particles cluster
around $\theta=0$ and form a Dirac peak for $T=0$. For $T\rightarrow
0$, they will remain localized around $\theta=0$. Thus, we can expand
Eq.  (\ref{in5}) to first order in
$\theta$. It becomes the equation  of a
harmonic oscillator
\begin{eqnarray}
\label{in8}
\frac{d^{2}\theta}{dt^{2}}+\omega^{2}\theta=0,
\end{eqnarray}
and the equations of motion are explicitly given by
\begin{eqnarray}
\label{in9}
\theta(t)=\frac{v_{0}}{\omega}\sin(\omega t)+\theta_{0}\cos(\omega t),
\end{eqnarray}
\begin{eqnarray}
\label{in10}
v(t)=v_{0}\cos(\omega t)-\theta_{0}\omega\sin(\omega t).
\end{eqnarray}
For $T\ll T_{c}$, the parameter $B$ is given by
\begin{eqnarray}
\label{in11}
\frac{B}{B_{max}}=\pm \left (1-\frac{T}{4T_{c}}\right )
\end{eqnarray}
where $B_{max}=2T_{c}=kM/2\pi$ \cite{vatt}. Therefore, to leading order, we find that the pulsation of the particle trajectory is 
\begin{eqnarray}
\label{in12}
\omega^{2}=\frac{kM}{2\pi}.
\end{eqnarray}

We shall not attempt in this paper to study the full kinetic equation
(\ref{in4}). We shall limit ourselves to analytically study the
behaviour of certain quantities that enter in this equation. To
emphasize the physical structure of the kinetic equation (\ref{in4}),
we introduce two functions
\begin{eqnarray}
\label{in13}
{\cal C}(t,\tau)=\int d\theta_{1}dv_{1}{\cal F}(1\rightarrow 0,t){\cal F}(1\rightarrow 0,t-\tau)\rho(\theta_{1})f(v_{1}),\nonumber\\
\end{eqnarray}
\begin{eqnarray}
\label{in14}
\Psi(t,\tau)=\beta\int d\theta_{1}dv_{1}{\cal F}(1\rightarrow 0,t)\qquad\qquad\qquad\nonumber\\
\times {\cal F}(1\rightarrow 0,t-\tau)v_{1}(t-\tau)\rho(\theta_{1})f(v_{1}),
\end{eqnarray}
in terms of which Eq. (\ref{in4}) can be rewritten
\begin{eqnarray}
\label{in15}
\frac{\partial P}{\partial t}+v\frac{\partial P}{\partial \theta}+\langle F\rangle \frac{\partial P}{\partial v}=\frac{\partial}{\partial v}\int_{0}^{t}d\tau  \biggl\lbrace  {\cal C}(t,\tau)\frac{\partial }{\partial v}\nonumber\\
-\Psi(t,\tau)\biggr\rbrace  P(\theta(t-\tau),v(t-\tau),t-\tau).\qquad
\end{eqnarray}
The function ${\cal C}(t,\tau)$ which appears in the {\it diffusion
term} represents the force auto-correlation function. It will be
specifically studied in Sec. \ref{sec_auto}. On the other hand, the
function $\Psi (t,\tau)$ which appears in the {\it drift term} is
connected to the friction force which will be specifically studied in
Sec. \ref{sec_fric} (see in particular the quantity $I$). Then, the
difficulty with Eq. (\ref{in15}) lies in that fact that it is non
Markovian: we need to know the value of $P(\theta,v,t)$ at previous
times to pass from $t$ to $t+dt$. In Sec.  \ref{sec_time}, in order to
get some insight into the problem, we shall consider a simpler kinetic
equation where we neglect non-Markovian terms altogether but keep the
time dependence of the diffusion and friction coefficients.

\section{The force auto-correlation function}
\label{sec_auto}

\subsection{The temporal correlation function}
\label{sec_autotemp}

One quantity of great interest in the kinetic theory is the force
auto-correlation function. Indeed, in ordinary circumstances, the
diffusion coefficient in the Fokker-Planck equation is expressed as
the time integral of the auto-correlation function through the Kubo
formula.  The temporal auto-correlation of the fluctuating force can
be decomposed into
\begin{eqnarray}
\label{auto1}
\langle {\cal F}(0){\cal F}(t)\rangle=\langle F(0)F(t)\rangle-\langle F(0)\rangle \langle F(t)\rangle,
\end{eqnarray}
where ${\cal F}(t)=F(t)-\langle F(t)\rangle$ is the total fluctuating force acting on the test particle at time $t$. Using $F(t)=\sum_{i}F(i\rightarrow 0,t)$, we get
\begin{eqnarray}
\label{auto2}
\langle {F}(0){F}(t)\rangle=\sum_{i,j}\langle F(i\rightarrow 0,0)F(j\rightarrow 0,t)\rangle\nonumber\\
=\sum_{i}\langle F(i\rightarrow 0,0)F(i\rightarrow 0,t)\rangle\nonumber\\
+\sum_{i\neq j}\langle F(i\rightarrow 0,0)F(j\rightarrow 0,t)\rangle.
\end{eqnarray}
Since the $N$-body distribution of the bath is a product of $N$
one-body distributions (see, e.g., \cite{vatt}) and since the particles
are identical, we obtain
\begin{eqnarray}
\label{auto3}
\langle {F}(0){F}(t)\rangle=
N\langle F(1\rightarrow 0,0)F(1\rightarrow 0,t)\rangle\nonumber\\
+N(N-1)\langle F(1\rightarrow 0,0)\rangle\langle F(1\rightarrow 0,t)\rangle.
\end{eqnarray}
Accounting that $\langle F(t)\rangle=N\langle F(1\rightarrow 0,t)\rangle$, we 
get
\begin{eqnarray}
\label{auto4}
\langle {\cal F}(0){\cal F}(t)\rangle=N\langle F(1\rightarrow 0,0)F(1\rightarrow 0,t)\rangle\nonumber\\
-N\langle F(1\rightarrow 0,0)\rangle\langle F(1\rightarrow 0,t)\rangle,
\end{eqnarray} 
which can be written
\begin{eqnarray}
\label{auto5}
\langle {\cal F}(0){\cal F}(t)\rangle=N\langle {\cal F}(1\rightarrow 0,0){\cal F}(1\rightarrow 0,t)\rangle,
\end{eqnarray} 
where ${\cal F}(1\rightarrow 0,t)=F(1\rightarrow 0,t)-\langle
F(1\rightarrow 0,t)\rangle$ is the fluctuating force produced by
particle $1$ on the test particle. Explicitly,
\begin{eqnarray}
\label{auto6}
\langle {\cal F}(0){\cal F}(t)\rangle=\int {\cal F}(1\rightarrow 0,0){\cal F}(1\rightarrow 0,t)\rho(\theta_{1})f(v_{1})d\theta_{1}dv_{1}.\nonumber\\
\end{eqnarray} 
We note that this combination of terms enters in the diffusion term in
Eq. (\ref{in4}); this is precisely the function ${\cal C}(t,\tau)$
defined in Eq. (\ref{in13}) where we have taken the origin of times
at $t=0$. Let us first compute the quantity
\begin{eqnarray}
\label{auto7}
\lbrace {F}(0){F}(t)\rbrace=N\langle F(1\rightarrow 0,0)F(1\rightarrow 0,t)\rangle.
\end{eqnarray}
Explicitly, we have
\begin{eqnarray}
\label{auto8}
\lbrace {F}(0){F}(t)\rbrace=\int F(1\rightarrow 0,0)F(1\rightarrow 0,t)\nonumber\\
\times \rho(\theta_{1})f(v_{1})d\theta_{1}dv_{1}.
\end{eqnarray}
Using Eq. (\ref{in3}), we get
\begin{eqnarray}
\label{auto9}
\lbrace {F}(0){F}(t)\rbrace=\frac{k^{2}}{4\pi^{2}}\int \sin(\theta-\theta_{1}) \sin(\theta(t)-\theta_{1}(t))\nonumber\\
\times \rho(\theta_{1})f(v_{1})d\theta_{1}dv_{1},
\end{eqnarray}
where $\theta_{i}(t)$ denotes the position at time $t$ of the $i$-th particle located at $\theta_{i}$ at $t=0$. Now, using the equation of motion (\ref{in9}), we obtain
\begin{eqnarray}
\label{auto10}
\theta(t)-\theta_{1}(t)=\frac{u}{\omega}\sin(\omega t)+\phi \cos(\omega t),
\end{eqnarray}
where $\phi=\theta-\theta_{1}$ and $u=v-v_{1}$. Substituting these results in Eq. (\ref{auto9}) we get
\begin{eqnarray}
\label{auto11}
\lbrace {F}(0){F}(t)\rbrace=\frac{k^{2}}{4\pi^{2}}\int dv_{1} f(v_{1})\int d\theta_{1} \rho(\theta_{1}) \sin\phi \nonumber\\
\times  \sin\left\lbrack \frac{u}{\omega}\sin(\omega t)+\phi\cos(\omega t)\right\rbrack. 
\end{eqnarray}
In the $T\rightarrow 0$ limit, the spatial distribution of the particles (\ref{stat3}) can be approximated by
\begin{eqnarray}
\label{auto12}
\rho(\theta)=M\omega\left (\frac{\beta}{2\pi}\right )^{1/2}e^{-\beta \omega^{2}\frac{\theta^{2}}{2}}.
\end{eqnarray}
Using furthermore $\sin(x)\simeq x$ in Eq. (\ref{auto11}) for small $\theta$, $v$ and performing the Gaussian integrations, we finally obtain
\begin{eqnarray}
\label{auto13}
\lbrace {F}(0){F}(t)\rbrace=\frac{Mk^{2}}{4\pi^{2}}\frac{v\theta}{\omega}\sin(\omega t) 
+\frac{Mk^{2}}{4\pi^{2}}\left (\theta^{2}+\frac{1}{\beta\omega^{2}}\right )\cos(\omega t). \nonumber\\
\end{eqnarray}
In the preceding expansions, we have implicitly assumed that the
coordinates of the field particles {\it and} of the test particle scale as
$v\sim \sqrt{2/\beta}$ and $\theta\sim
(1/\omega)\sqrt{2/\beta}$. Therefore, our asymptotic expansion is
valid to order $T$ for $T\rightarrow 0$. 

Now, the correlation function of the fluctuating force is given by
\begin{eqnarray}
\label{auto14}
\langle {\cal F}(0){\cal F}(t)\rangle =\lbrace {F}(0){F}(t)\rbrace-N\langle F(1\rightarrow 0,0)\rangle\langle F(1\rightarrow 0,t)\rangle,\nonumber\\ 
\end{eqnarray}
where
\begin{eqnarray}
\label{auto15}
\langle {F}(1\rightarrow 0,t)\rangle=-\frac{k}{2\pi N}\int \sin(\theta(t)-\theta_{1}(t))\rho(\theta_{1})f(v_{1})d\theta_{1}dv_{1}. \nonumber\\
\end{eqnarray}
Using the same approximations as before, we obtain
\begin{eqnarray}
\label{auto16}
\langle {F}(1\rightarrow 0,t)\rangle=-\frac{k}{2\pi }\left \lbrack \frac{v}{\omega}\sin(\omega t)+\theta\cos(\omega t)\right \rbrack.
\end{eqnarray}
Combining the previous results, we  get
\begin{eqnarray}
\label{auto17}
\langle {\cal F}(0){\cal F}(t)\rangle =\frac{k^{2}}{4\pi^{2}}\frac{M}{\beta\omega^{2}}\cos(\omega t).
\end{eqnarray}
Using Eq. (\ref{in12}), we finally obtain
\begin{eqnarray}
\label{auto18}
\langle {\cal F}(0){\cal F}(t)\rangle =\frac{k}{2\pi\beta}\cos(\omega t)+O(T^{2}).
\end{eqnarray}
We note that, to order $T$, the correlation function of the
fluctuating force depends only on the ellapsed time $t$ and not on the
initial instant $t=0$. We also note that the correlation function is
periodic with the same pulsation $\omega$ as the particle trajectory
and that it averages to zero over a period.

\subsection{The diffusion coefficient}
\label{sec_difc}

\begin{figure}
\centering
\includegraphics[width=8cm]{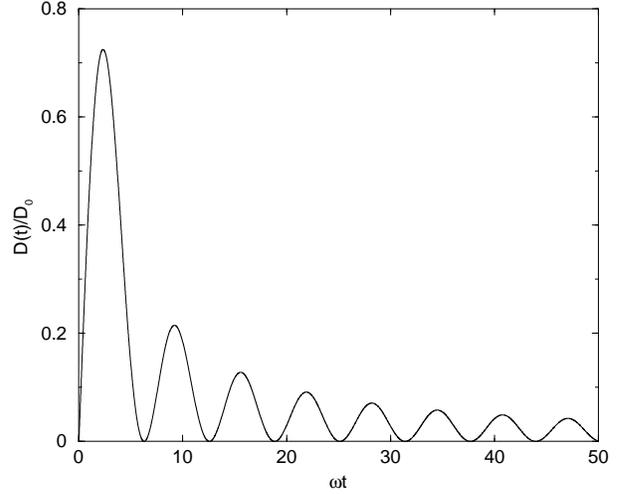}
\caption{The function $D(t)/D_{0}$ where $D_{0}=\frac{k^{2}}{4\pi^{2}}\frac{M}{\beta\omega^{3}}$. }
\label{diff}
\end{figure}

If $\Delta v=\int_{0}^{t}{\cal F}(t')dt'$ denotes the increment of
velocity of the test particle caused by the fluctuating force during an
interval of time $t$, we define the diffusion coefficient by
\begin{eqnarray}
\label{auto19}
D(t)=\frac{\langle (\Delta v)^{2}\rangle}{2t}.
\end{eqnarray} 
This can be rewritten
\begin{eqnarray}
\label{auto20}
D(t)=\frac{1}{2t}\int_{0}^{t}dt'\int_{0}^{t}dt'' \langle {\cal F}(t'){\cal F}(t'')\rangle.
\end{eqnarray} 
Since the correlation function depends only on the time interval $|t''-t'|$, we also have
\begin{eqnarray}
\label{auto21}
D(t)=\frac{1}{t}\int_{0}^{t}dt'\int_{t'}^{t}dt'' \langle {\cal F}(0){\cal F}(t''-t')\rangle.
\end{eqnarray} 
Setting $\tau=t''-t'$, we get
\begin{eqnarray}
\label{auto22}
D(t)=\frac{1}{t}\int_{0}^{t}dt'\int_{0}^{t-t'}d\tau \langle {\cal F}(0){\cal F}(\tau)\rangle,
\end{eqnarray}
or, equivalently,
\begin{eqnarray}
\label{auto23}
D(t)=\frac{1}{t}\int_{0}^{t}d\tau\int_{0}^{t-\tau} dt' \langle {\cal F}(0){\cal F}(\tau)\rangle.
\end{eqnarray}
Finally, we obtain
\begin{eqnarray}
\label{auto24}
D(t)=\frac{1}{t}\int_{0}^{t}d\tau (t-\tau) \langle {\cal F}(0){\cal F}(\tau)\rangle.
\end{eqnarray}
If the correlation function decreases sufficiently rapidly with time, taking $t\rightarrow +\infty$, we obtain the Kubo formula
\begin{eqnarray}
\label{auto25}
D=\int_{0}^{+\infty} \langle {\cal F}(0){\cal F}(\tau)\rangle d\tau.
\end{eqnarray}
However, since in the present situation the temporal correlation
function has an oscillatory behaviour, this formula is not
applicable. According to Eq. (\ref{auto24}), the diffusion coefficient
can be written $D=D_{1}-D_{2}$ where
\begin{eqnarray}
\label{auto26}
D_{1}=\int_{0}^{t} \langle {\cal F}(0){\cal F}(\tau)\rangle d\tau,
\end{eqnarray}
and
\begin{eqnarray}
\label{auto27}
D_{2}=\frac{1}{t}\int_{0}^{t} \langle {\cal F}(0){\cal F}(\tau)\rangle \tau d\tau.
\end{eqnarray}
Using Eq. (\ref{auto17}), we find that
\begin{eqnarray}
\label{auto28}
D_{1}=\frac{k^{2}}{4\pi^{2}}\frac{M}{\beta\omega^{3}}\sin(\omega t),
\end{eqnarray}
\begin{eqnarray}
\label{auto29}
D_{2}=\frac{k^{2}}{4\pi^{2}}\frac{M}{\beta\omega^{3}}\biggl\lbrack \sin(\omega t)+\frac{1}{\omega t}\cos(\omega t)-\frac{1}{\omega t}\biggr\rbrack.
\end{eqnarray}
Thus, we obtain
\begin{eqnarray}
\label{auto30}
D(t)=\frac{k^{2}}{4\pi^{2}}\frac{M}{\beta\omega^{3}}\frac{1-\cos(\omega t)}{\omega t}. 
\end{eqnarray}
The diffusion coefficient is periodic and goes to zero at each time $t_{n}=(2\pi/\omega)n$ with $n=1,2,...$ (see Fig. \ref{diff}). For $t\rightarrow 0$, it behaves like
\begin{eqnarray}
\label{auto31}
D(t)=\frac{k^{2}}{8\pi^{2}}\frac{M}{\beta\omega^{2}}t=\frac{k}{4\pi\beta}t.
\end{eqnarray}
On the other hand, $D(t)\rightarrow 0$ for $t\rightarrow +\infty$.

\subsection{The spatial correlation function}
\label{sec_autospat}

Let us finally provide the exact expression of the spatial correlation
function $\langle {\cal F}(\theta){\cal F}(\theta')\rangle$ in the
case where the correlations between particles are neglected (as
before).  The effect of correlations is considered in \cite{vatt,bh}
for the homogeneous phase. The case of the inhomogeneous phase will be
considered elsewhere.

Repeating the same steps as in Sec. \ref{sec_autotemp}, the spatial correlations of the fluctuating force can be written
\begin{eqnarray}
\label{sf1}
\langle {\cal F}(\theta){\cal F}(\theta')\rangle =\lbrace F(\theta)F(\theta')\rbrace-\frac{1}{N}\langle F(\theta)\rangle \langle F(\theta')\rangle,
\end{eqnarray} 
where 
\begin{eqnarray}
\label{sf2}
\lbrace F(\theta)F(\theta')\rbrace=\frac{k^{2}}{4\pi^{2}}\int_{0}^{2\pi} \sin(\theta-\theta_{1})\sin(\theta'-\theta_{1})\rho(\theta_{1})d\theta_{1},\nonumber\\
\end{eqnarray} 
and 
\begin{eqnarray}
\label{sf3}
\langle F(\theta)\rangle=B\sin\theta=\frac{kM}{2\pi}\frac{I_{1}(\beta B)}{I_{0}(\beta B)}\sin\theta.
\end{eqnarray} 
Using Eq. (\ref{stat3}), we have 
\begin{eqnarray}
\label{sf4}
\lbrace F(\theta)F(\theta')\rbrace=\frac{k^{2}M}{8\pi^{3}I_{0}(\beta B)}\qquad\qquad\qquad\qquad\nonumber\\
\times\int_{0}^{2\pi} \sin(\theta-\theta_{1})\sin(\theta'-\theta_{1})e^{-\beta B\cos\theta_{1}}d\theta_{1}.
\end{eqnarray} 
Expanding the trigonometric functions and using  the identities
\begin{eqnarray}
\label{sf5}
\int_{0}^{2\pi}\sin\theta_{1}\cos\theta_{1}e^{-\beta B\cos\theta_{1}}d\theta_{1}=0,
\end{eqnarray} 
\begin{eqnarray}
\label{sf6}
\int_{0}^{2\pi}\sin^{2}\theta_{1}e^{-\beta B\cos\theta_{1}}d\theta_{1}=\frac{2\pi}{\beta B}I_{1}(\beta B),
\end{eqnarray}
the integrals in Eq. (\ref{sf4}) can be easily  performed. Then, using Eqs. (\ref{sf3}) and (\ref{sf1}), we finally obtain
\begin{eqnarray}
\label{sf7}
\langle {\cal F}(\theta){\cal F}(\theta')\rangle =\frac{k^{2}M}{4\pi^{2}}\biggl\lbrack \frac{I_{1}(x)}{xI_{0}(x)}\cos(\theta-\theta')\nonumber\\
+\biggl (1-\frac{I_{1}(x)^{2}}{I_{0}(x)^{2}}-\frac{2I_{1}(x)}{xI_{0}(x)}\biggr )\sin\theta\sin\theta'\biggr \rbrack,
\end{eqnarray} 
where we have set $x=\beta B$. We note that, due to the inhomogeneity
of the system, the correlation function of the fluctuating force is
not a function of $|\theta-\theta'|$ alone.

\begin{figure}
\centering
\includegraphics[width=8cm]{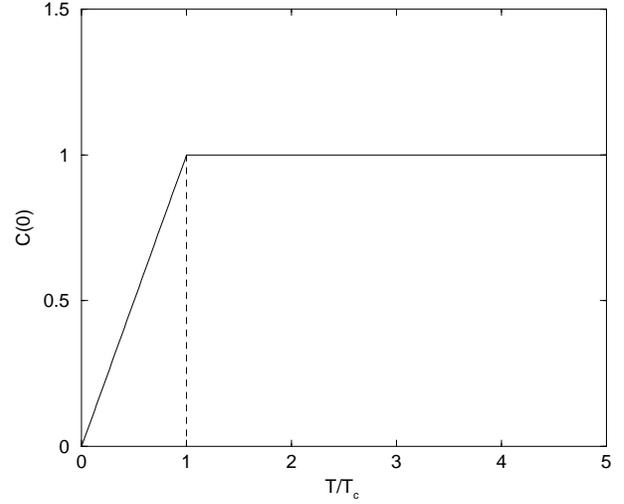}
\caption{The normalized variance of the fluctuating force $C(0)=\frac{8\pi^{2}}{k^{2}M}\langle {\cal F}(0)^{2}\rangle$ at $\theta=0$ as a function of the temperature when the correlations between particles are neglected. }
\label{auto}
\end{figure}

Let us consider particular cases. If we take $\theta'=\theta$, we see that $\langle {\cal F}(\theta)^{2}\rangle$ depends on $\theta$ through a term $\sin^{2}\theta$. If we take $\theta=0$ and $\theta'=\phi$, we obtain
\begin{eqnarray}
\label{sf8}
\langle {\cal F}(0){\cal F}(\phi)\rangle =\frac{k^{2}M}{4\pi^{2}}\frac{I_{1}(x)}{xI_{0}(x)}\cos\phi.
\end{eqnarray}
For $T\ge T_{c}$ (homogeneous phase), we have $x=0$ and Eq. (\ref{sf8}) reduces to
\begin{eqnarray}
\label{sf9}
\langle {F}(0){F}(\phi)\rangle =\frac{k^{2}M}{8\pi^{2}}\cos\phi.
\end{eqnarray}
We recover the result of \cite{vatt,bh} when the correlations between
particles are neglected. For $T<T_{c}$, using Eq. (\ref{stat6}), we
find that
\begin{eqnarray}
\label{sf10}
\langle {\cal F}(0){\cal F}(\phi)\rangle =\frac{kT}{2\pi}\cos\phi.
\end{eqnarray}
The dependence of this correlation function with the temperature is
plotted in Fig. \ref{auto}.  Coming back to the function (\ref{sf7}),
and considering the homogeneous phase $x=0$, we get
\begin{eqnarray}
\label{sf11}
\langle {\cal F}(\theta){\cal F}(\theta')\rangle =\frac{k^{2}M}{4\pi^{2}}\cos (\theta-\theta').
\end{eqnarray}
On the other hand, for $T\rightarrow 0$ corresponding to $x\rightarrow +\infty$, we find that
\begin{eqnarray}
\label{sf12}
\langle {\cal F}(\theta){\cal F}(\theta')\rangle =\frac{kT}{2\pi}\cos (\theta+\theta')+O(T^{2}).
\end{eqnarray}
In particular, 
\begin{eqnarray}
\label{sf13}
\langle {\cal F}(\theta)^{2}\rangle =\frac{kT}{2\pi}\cos (2\theta)+O(T^{2}).
\end{eqnarray}
If we assume that $\theta\sim \sqrt{T}$ as in Sec. \ref{sec_autotemp}, then 
\begin{eqnarray}
\label{sf14}
\langle {\cal F}^{2}\rangle =\frac{kT}{2\pi}+O(T^{2}),
\end{eqnarray}
which coincides with Eq. (\ref{auto18}) when $t=0$.

\section{The friction force}
\label{sec_fric}

We shall now compute the frictional force 
\begin{eqnarray}
\label{fric0}
\langle F_{\rm fr}(t)\rangle=\beta\int_{0}^{t}d\tau\int d\theta_{1}dv_{1}{\cal F}(1\rightarrow 0,t){\cal F}(0\rightarrow 1,t-\tau)\nonumber\\
\times v_{1}(t-\tau) \rho(\theta_{1})f(v_{1}),\qquad 
\end{eqnarray}
experienced by a test particle with {\it prescribed} trajectory given
by Eq. (\ref{in9}). The expression (\ref{fric0}) can be directly
obtained from a linear response theory as done by Kandrup
\cite{kandrup2} for the gravitational interaction. The friction arises as
the response of the field particles to the perturbation caused by the
test particle as in a polarization process. Note that when the test
particle is described by a distribution function $P(\theta,v,t)$
[instead of having a prescribed trajectory] as in Sec. \ref{sec_in},
the combination of terms (\ref{fric0}) also enters in the friction
term of equation (\ref{in4}) although the expression is more
complicated as it involves the history of the distribution function of
the test particle. 

Since the fluctuating force averages to zero, Eq. (\ref{fric0}) can
also be written as
\begin{eqnarray}
\label{fric0b}
\langle F_{\rm fr}(t)\rangle=\beta\int_{0}^{t}d\tau\int d\theta_{1}dv_{1}{F}(1\rightarrow 0,t){\cal F}(0\rightarrow 1,t-\tau)\nonumber\\
\times v_{1}(t-\tau) \rho(\theta_{1})f(v_{1}).\qquad 
\end{eqnarray}
Taking the origin of times at $t=0$, we shall first compute the quantity
\begin{eqnarray}
\label{fric1}
I(t)=\int F(1\rightarrow 0,0)F(1\rightarrow 0,t)v_{1}(t)\nonumber\\
\times \rho(\theta_{1})f(v_{1})d\theta_{1}dv_{1},
\end{eqnarray}
which is related to the function $\Psi(t,\tau)$ defined in
Eq. (\ref{in14}).  Using the fact that
\begin{eqnarray}
\label{fric2}
v_{1}(t)=v_{1}\cos(\omega t)-\theta_{1}\omega \sin(\omega t),
\end{eqnarray}
we can write $I=I_{1}+I_{2}$ where
\begin{eqnarray}
\label{fric3}
I_{1}(t)=\cos(\omega t)\int F(1\rightarrow 0,0)F(1\rightarrow 0,t)v_{1}\nonumber\\
\times \rho(\theta_{1})f(v_{1})d\theta_{1}dv_{1},
\end{eqnarray}
\begin{eqnarray}
\label{fric4}
I_{2}(t)=-\omega\sin(\omega t)\int F(1\rightarrow 0,0)F(1\rightarrow 0,t)\theta_{1}\nonumber\\
\times \rho(\theta_{1})f(v_{1})d\theta_{1}dv_{1}.
\end{eqnarray}

The first integral can be rewritten
\begin{eqnarray}
\label{fric5}
I_{1}(t)=\frac{k^{2}}{4\pi^{2}}\cos(\omega t)\int dv_{1} f(v_{1})v_{1}\int d\theta_{1} \rho(\theta_{1}) \sin\phi \nonumber\\
\times  \sin\left\lbrack \frac{u}{\omega}\sin(\omega t)+\phi\cos(\omega t)\right\rbrack.
\end{eqnarray}
Using the same approximations as in Sec. \ref{sec_autotemp}, we obtain
\begin{eqnarray}
\label{fric6}
I_{1}(t)=-\frac{k^{2}}{4\pi^{2}}\frac{M\theta}{2\omega\beta}\sin (2\omega t).
\end{eqnarray}
Our asymptotic expansion is here valid to order $T^{3/2}$.
The second integral can be rewritten
\begin{eqnarray}
\label{fric7}
I_{2}(t)=-\frac{k^{2}}{4\pi^{2}}\omega \sin(\omega t)\int dv_{1} f(v_{1})\int d\theta_{1} \rho(\theta_{1}) \theta_{1}\sin\phi \nonumber\\
\times  \sin\left\lbrack \frac{u}{\omega}\sin(\omega t)+\phi\cos(\omega t)\right\rbrack,\quad
\end{eqnarray}
leading to
\begin{eqnarray}
\label{fric8}
I_{2}(t)=\frac{k^{2}}{4\pi^{2}}\frac{Mv}{\beta\omega^{2}}\sin^{2}(\omega t)+\frac{k^{2}}{4\pi^{2}}\frac{M\theta}{\beta\omega}\sin(2\omega t).
\end{eqnarray}
Summing these results, we get
\begin{eqnarray}
\label{fric9}
I(t)=\frac{k^{2}}{4\pi^{2}}\frac{Mv}{\beta\omega^{2}}\sin^{2}(\omega t)+\frac{k^{2}}{8\pi^{2}}\frac{M\theta}{\beta\omega}\sin(2\omega t).
\end{eqnarray}
Since $\sin^{2}(\omega t)=\lbrack 1-\cos(2\omega t)\rbrack/2$, the function $I(t)$ is periodic with pulsation $2\omega$, i.e. twice the pulsation of the orbiting particles. However, this function also contains a constant component which does not average to zero over a period. We have instead
\begin{eqnarray}
\label{fric9b}
\langle I(t)\rangle=\frac{1}{T}\int_{0}^{T}I(t)dt=\frac{k^{2}Mv}{8\pi^{2}\beta \omega^{2}},
\end{eqnarray}
where $T$ here denotes the period. This implies that {\it the effects of the
friction accumulate with time}.  

\begin{figure}
\centering
\includegraphics[width=8cm]{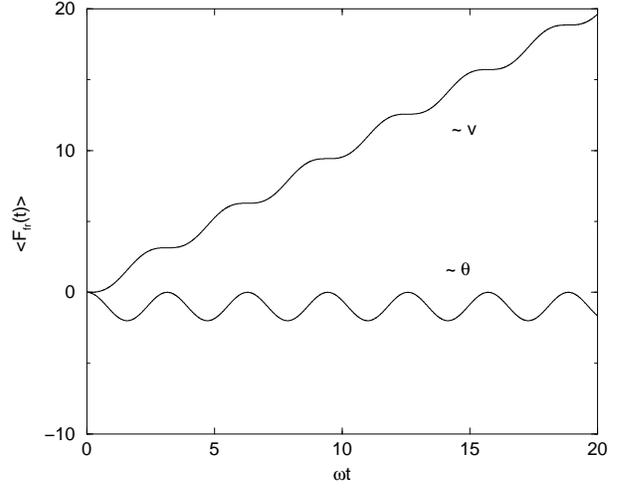}
\caption{The temporal behaviour of the components of the friction force proportional to $v$ and $\theta$. }
\label{fric}
\end{figure}

To obtain the complete friction force, we also have to calculate the quantity
\begin{eqnarray}
\label{fric10}
J(t)=\int F(1\rightarrow 0,0)\langle F(1\rightarrow 0,t)\rangle v_{1}(t)\nonumber\\
\times \rho(\theta_{1})f(v_{1})d\theta_{1}dv_{1}.
\end{eqnarray}
Using 
\begin{eqnarray}
\label{fric11}
\langle {F}(0\rightarrow 1,t)\rangle=-\frac{k}{2\pi }\left \lbrack \frac{v_{1}}{\omega}\sin(\omega t)+\theta_{1}\cos(\omega t)\right \rbrack,
\end{eqnarray}
and performing the same approximations as before, we find that
\begin{eqnarray}
\label{fric12}
J(t)=\frac{k^{2}}{4\pi^{2}}\int \phi \left (\frac{v_{1}}{\omega}\sin\omega t+\theta_{1}\cos\omega t\right )\nonumber\\
\times \left (v_{1}\cos\omega t-\theta_{1}\omega \sin\omega t\right )\rho(\theta_{1})f(v_{1})d\theta_{1}dv_{1}.
\end{eqnarray}
By using the parity of the velocity and angular distributions, we get
\begin{eqnarray}
\label{fric13}
J(t)=\frac{k^{2}\theta}{8\pi^{2}}\sin (2\omega t)\int \left (\frac{v_{1}^{2}}{\omega}-\theta_{1}^{2}\omega\right )\rho(\theta_{1})f(v_{1})d\theta_{1}dv_{1}.\nonumber\\
\end{eqnarray} 
Finally, performing the Gaussian integrations, we find that $J=0$. Therefore, the frictional force at time $t$ is simply given by 
\begin{eqnarray}
\label{fric14}
\langle F_{\rm fr}(t)\rangle=-\beta\int_{0}^{t}I(-\tau)d\tau.
\end{eqnarray}
Focusing on the component proportional to the velocity 
of the test particle (which contains the diverging contribution for $t\rightarrow +\infty$), we get
\begin{eqnarray}
\label{fric15}
\langle F_{\rm fr}(t)\rangle=-\frac{k^{2}M}{4\pi^{2}}\frac{v(t)}{\omega^{2}}\int_{0}^{t}\sin^{2}( \omega\tau) d\tau\equiv -\xi(t) v. 
\end{eqnarray}
We note that, in the present situation, the friction coefficient is  {\it not} given by an Einstein relation. Keeping only the diverging contribution coming from the non-vanishing averaged value of $I(t)$ given by Eq. (\ref{fric9b}), we find that 
\begin{eqnarray}
\label{fric16}
\xi(t)=\frac{k^{2}M}{8\pi^{2}}\frac{1}{\omega^{2}}t=\frac{kt}{4\pi},
\end{eqnarray}
so that
\begin{eqnarray}
\label{fric17}
\langle F_{\rm fr}(t)\rangle=-\frac{kt}{4\pi} v(t). 
\end{eqnarray}
Therefore, the friction coefficient {\it increases linearly with time}. If we consider only that part (linear in $t$) of the friction coefficient, we find that it is related to the diffusion coefficient (\ref{auto30}) by 
\begin{eqnarray}
\label{fric18}
\xi(t)=\beta D(t) \frac{\omega^{2}t^{2}}{2(1-\cos\omega t)},
\end{eqnarray}
which resembles the Einstein relation. For $t\rightarrow 0$, we find
that $\xi(t)=D(t)\beta$.

If we now consider the full expression of the friction force, we get
\begin{eqnarray}
\label{fric19}
\langle F_{\rm fr}(t)\rangle=-\frac{k^{2}M}{8\pi^{2}}\frac{1}{\omega^{2}}\left\lbrack v\left (t-\frac{\sin 2\omega t}{2\omega}\right )+\frac{\theta}{2}(\cos 2\omega t-1)\right \rbrack.\nonumber\\
\end{eqnarray}
The temporal behaviours of the components of the friction force
proportional to $v$ and $\theta$ are represented in
Fig. \ref{fric}. For $t\rightarrow +\infty$, we recover
Eq. (\ref{fric17}) and for $t\rightarrow 0$, we have
\begin{eqnarray}
\label{fric20}
\langle F_{\rm fr}(t)\rangle=-\frac{k^{2}M}{4\pi^{2}}\left (\frac{v}{3}t^{3}-\frac{\theta}{2}t^{2}\right ).
\end{eqnarray}

Finally, with the quantities defined in Secs. \ref{sec_auto} and \ref{sec_fric}, the kinetic equation (\ref{in4}) can be written
\begin{eqnarray}
\label{fric23}
\frac{\partial P}{\partial t}+v\frac{\partial P}{\partial \theta}+\langle F\rangle \frac{\partial P}{\partial v}=\nonumber\\
\frac{\partial}{\partial v}\int_{0}^{t}d\tau
\biggl\lbrace \langle  {\cal F}(t){\cal F}(t-\tau)\rangle \frac{\partial }{\partial v}+\beta I(-\tau)\biggr\rbrace \nonumber\\
\times P(\theta(t-\tau),v(t-\tau),t-\tau).
\end{eqnarray}
Using Eq. (\ref{auto18}) and keeping only the constant term in Eq. (\ref{fric9}), we get
\begin{eqnarray}
\label{fric24}
\frac{\partial P}{\partial t}+v\frac{\partial P}{\partial \theta}+\langle F\rangle \frac{\partial P}{\partial v}=\frac{k^{2}M}{8\pi^{2}}\frac{1}{\beta\omega^{2}} \frac{\partial}{\partial v}\qquad\nonumber\\
\times\int_{0}^{t}d\tau
\biggl\lbrace 2\cos(\omega\tau)\frac{\partial }{\partial v}+\beta v(t)\biggr\rbrace  P(\theta(t-\tau),v(t-\tau),t-\tau).\nonumber\\
\end{eqnarray}
The study of these kinetic equations is beyond the scope of the
present paper. Let us simply state that the kinetic theory of
inhomogeneous systems with long-range interactions is far from being
completely understood. In particular, it is not clear whether
Eq. (\ref{fric24}), and more generally Eq. (\ref{in1}), satisfies an
$H$-theorem. Indeed, starting from the general kinetic equation (\ref{in1}), introducing the Boltzmann entropy $S=-\int f\ln f d\theta dv$ and using standard methods, we can 
put the rate of entropy production in the form
\begin{eqnarray}
\label{thH1}
\dot S=\frac{1}{2}\int d\theta dv d\theta_{1} dv_{1}\frac{1}{f f_{1}}\int_{0}^{t}d\tau Q(t)G(t,t-\tau) Q(t-\tau),\nonumber\\
\end{eqnarray}
where
\begin{eqnarray}
\label{thH2}
Q(t)\equiv \left\lbrack {\cal F}(1\rightarrow 0,t)\frac{\partial}{\partial v}+{\cal F}(0\rightarrow 1,t)\frac{\partial}{\partial v_{1}}\right \rbrack\nonumber\\
\times f(\theta,v,t)f(\theta_{1},v_{1},t).
\end{eqnarray}
In the course of the calculations, we have inverted the dummy
variables $\theta,v$ and $\theta_{1},v_{1}$ and taken the half sum of
the resulting expressions (see, e.g., \cite{pa}). We see that the
$H$-theorem $\dot S\ge 0$ is not granted. This depends on the temporal
correlations of $Q(t)$. Furthermore, it is not clear whether
Eq. (\ref{in1}) conserves energy and whether it converges towards some
form of equilibrium (Maxwellian or other) for large times. If we take
for granted that the system {\it must} converge towards statistical
equilibrium for $t\rightarrow +\infty$, this implies that the kinetic
theory may not be complete; one may have to relax certain simplifying
approximations and consider next order terms in the expansion in $1/N$
of the correlation functions. Alternatively, if we rely on the kinetic
theory to give the justification (or not) of the statistical
equilibrium state, one may conclude that the Boltzmannian distribution
may not be reached (at least in a strict sense) for inhomogeneous
systems.  There might be dynamical anomalies preventing the system
from reaching equilibrium. These are open questions left for future
works. Note, however, that the preceding discussion does not favour
other forms of entropy. The Boltzmann entropy remains the most
fundamental even if the $H$-theorem may not be rigorously valid.

\section{Kinetic equation with time dependent friction}
\label{sec_time}

\subsection{General solution}
\label{sec_gen}

The kinetic equation (\ref{fric24}) is complicated due to
non Markovian effects and spatial inhomogeneity encapsulated in the
advection term. In addition, one striking novel aspect of our study is
the time dependence of the diffusion coefficient and friction
force. Therefore, in a first attempt to investigate the effects
produced by such terms, we shall consider a simpler kinetic equation
where we neglect non-Markovian effects and advective terms altogether
but keep the time dependence of the diffusion and friction
coefficients. In that case, Eq. (\ref{fric24}) reduces to
\begin{eqnarray}
\label{time1}
\frac{\partial P}{\partial t}=\frac{\partial}{\partial v}\left (D(t)\frac{\partial P}{\partial v}+\xi(t)Pv\right ),
\end{eqnarray}
where $D(t)$ and $\xi(t)$ are given by Eqs. (\ref{auto26}) and
(\ref{fric16}). More generally, we shall consider equations of the
form (\ref{time1}) for arbitrary functions of time $D(t)$ and $\xi(t)$. An
interesting aspect of the problem is that such equations can be solved
analytically. We note that when $D$ and $\xi$ are constant, we get the
familiar Kramers equation. Its stationary solution is the Maxwellian
$P_{e}=Ae^{-\beta v^{2}/2}$ provided that $D$ and $\xi$ are related to
each other by the Einstein relation $\xi=D\beta$. We shall now
consider the case where $D$ and $\xi$ depend on time. By redefining
time such that $dt'=D(t)dt$, we are led to consider, without loss of
generality, equations of the form
\begin{eqnarray}
\label{time2}
\frac{\partial P}{\partial t}=\frac{\partial}{\partial v}\left (D\frac{\partial P}{\partial v}+\gamma h(t) Pv\right ),
\end{eqnarray} 
where $D$ and $\gamma$ are constant parameters.

Taking the Fourier transform of Eq. (\ref{time2}) with the conventions
\begin{eqnarray}
P(v)=\int \hat{P}(\xi)e^{i\xi v}d\xi,\qquad \hat{P}(\xi)=\int {P}(v)e^{-i\xi v}{dv\over 2\pi},\nonumber\\
\label{time3}
\end{eqnarray}
and using the relation
\begin{eqnarray}
P(v) v=\int v \hat{P}(\xi) e^{i\xi v}d\xi\qquad\qquad\nonumber\\
=\frac{1}{i}\int \hat{P}(\xi){\partial\over\partial \xi}(e^{i\xi v})d\xi=-\frac{1}{i}\int {\partial \hat{P}\over\partial\xi}e^{i\xi v}d\xi,\label{time4}
\end{eqnarray}
we get
\begin{eqnarray}
{\partial\hat{P}\over\partial t}=-D\xi^{2}\hat{P}-\gamma h(t)\xi {\partial\hat{P}\over\partial\xi}.\label{time5}
\end{eqnarray}
We introduce the change of variables
\begin{eqnarray}
f(y,t)=\hat{P}(H_{1}(t)y,t), \qquad \xi=H_{1}(t)y\label{time6}
\end{eqnarray}
and choose the function $H_{1}(t)$ such that
\begin{eqnarray}
{{\dot H}_{1}\over H_{1}}=\gamma h(t).\label{time7}
\end{eqnarray}
Substituting Eq. (\ref{time6}) in Eq. (\ref{time5}), we find that $f(y,t)$ satisfies
\begin{eqnarray}
{\partial f\over\partial t}+DH_{1}^{2}(t)y^{2}f=0.
\label{time8}
\end{eqnarray}
Let $H(t)$ be the primitive of $h(t)$ such that
\begin{eqnarray}
H(t)=\int_{0}^{t}h(\tau)d\tau.\label{time9}
\end{eqnarray}
Then, we choose $H_{1}$, solution of Eq. (\ref{time7}), such that
\begin{eqnarray}
H_{1}(t)=e^{\gamma H(t)}.\label{time10}
\end{eqnarray}
By convention, $H(0)=0$ and $H_{1}(0)=1$. Equation (\ref{time8}) can be integrated leading to
\begin{eqnarray}
f(y,t)=f(y,0)e^{-H_{2}(t)y^{2}},\label{time11}
\end{eqnarray}
where we have defined
\begin{eqnarray}
H_{2}(t)=D\int_{0}^{t} H_{1}(\tau)^{2}d\tau.
\label{time12}
\end{eqnarray}
Returning to original variables, we obtain
\begin{eqnarray}
\hat{P}(\xi,t)=\hat{P}_{0}\biggl ({\xi\over H_{1}(t)}\biggr )e^{-\chi^{2}(t)\xi^{2}},
\label{time13}
\end{eqnarray}
where we have defined
\begin{eqnarray}
\chi^{2}(t)\equiv \frac{H_{2}(t)}{H_{1}(t)^{2}}=D\int_{0}^{t}
e^{2\gamma\lbrack H(\tau)-H(t)\rbrack}d\tau.
\label{time14}
\end{eqnarray}
Defining
\begin{eqnarray}
q(v)=P_{0}(H_{1}(t)v)\quad\leftrightarrow \quad \hat{q}(\xi)={1\over H_{1}(t)}\hat{P}_{0}\biggl({\xi\over H_{1}(t)}\biggr )\nonumber\\
\end{eqnarray}
\begin{eqnarray}
g(v)=G(v/2\chi(t))\quad\leftrightarrow \quad
\hat{g}(\xi)=2\chi(t)\hat{G}(2\chi(t)\xi) \nonumber\\
\end{eqnarray}
where
\begin{eqnarray}
G(z)=e^{-z^{2}}\quad\leftrightarrow \quad \hat{G}(\xi)={1\over 2\sqrt{\pi}}e^{-\xi^{2}/4}
\end{eqnarray}
we can rewrite Eq. (\ref{time13}) in the form
\begin{eqnarray}
\hat{P}(\xi,t)=\sqrt{\pi}\frac{H_{1}(t)}{\chi(t)}\hat{q}(\xi)\hat{g}(\xi). \label{time15}
\end{eqnarray}
Taking the inverse Fourier transform, we can express the solution of Eq. (\ref{time2}) as a convolution
\begin{eqnarray}
{P}(v,t)=\sqrt{\pi}\frac{H_{1}(t)}{\chi(t)}\int
q(v-v')g(v'){dv'\over 2\pi}, \label{time16}
\end{eqnarray}
or, equivalently
\begin{eqnarray}
{P}(v,t)={H_{1}(t)\over
\sqrt{\pi}}\int_{-\infty}^{+\infty}e^{-x^{2}} P_{0}\lbrack H_{1}(t)(v-2\chi(t)x)\rbrack dx.\nonumber\\
\label{time17}
\end{eqnarray}
By direct substitution, we can check that Eq. (\ref{time17}) is indeed solution of Eq. (\ref{time2}). For $\gamma=0$ (pure diffusion) we find that 
\begin{eqnarray}
{P}(v,t)={1\over
\sqrt{\pi}}\int_{-\infty}^{+\infty}e^{-x^{2}} P_{0}(v-2x\sqrt{Dt})dx.
\label{time17new1}
\end{eqnarray}
For $D=0$, Eq. (\ref{time17}) reduces to
\begin{eqnarray}
{P}(v,t)={H_{1}(t)} P_{0}\lbrack H_{1}(t)v\rbrack.
\label{time17new2}
\end{eqnarray}
This can also be obtained by noticing that for $D=0$, Eq. (\ref{time2}) is an
equation of continuity. The equation of characteristic is
$dv/dt=-\gamma h(t)v$ which can be integrated into
$v(t)=v_{0}e^{-\gamma H(t)}$. Writing $P(v,t)dv=P_{0}(v_{0})dv_{0}$, we
finally get Eq. (\ref{time17new2}).

If $P_{0}(v)=\eta(v-v_0)$ is a step function with $\eta(v-v_0)=1$ for
$v<v_{0}$ and $\eta(v-v_0)=0$ for $v>v_{0}$, we find that
\begin{eqnarray}
{P}(v,t)=H_{1}(t)\Phi\biggl
({v-v_{0}/H_{1}(t)\over 2\chi(t)}\biggr ), \label{time18}
\end{eqnarray}
where
\begin{eqnarray}
\Phi(x)=\frac{1}{\sqrt{\pi}}\int_{x}^{+\infty} e^{-y^{2}}dy. 
\label{time19}
\end{eqnarray}
Alternatively, if $P_{0}(v)=\delta(v-v_{0})$, we obtain
\begin{eqnarray}
P(v,t)=\frac{1}{\sqrt{4\pi \chi(t)^{2}}}e^{-\frac{(v-v_{0}/H_{1}(t))^{2}}{4\chi(t)^{2}}}.
\label{time20}
\end{eqnarray}

\subsection{The case $h(t)=1$}
\label{sec_h1}

In the ordinary situation where the friction coefficient is constant ($h(t)=1$), we get
\begin{eqnarray}
H(t)=t, \quad H_{1}(t)=e^{\gamma t},  
\label{time21}
\end{eqnarray}
\begin{eqnarray}
H_{2}(t)=\frac{D}{2\gamma}(e^{2\gamma t}-1), \quad \chi^{2}(t)=\frac{D}{2\gamma}(1-e^{-2\gamma t}). \nonumber\\
\label{time22}
\end{eqnarray}
Equation (\ref{time18}) then takes the form
\begin{eqnarray}
P(v,t)=e^{\gamma t}\Phi\left (\frac{v-v_{0}e^{-\gamma t}}{\sqrt{\frac{2D}{\gamma}(1-e^{-2\gamma t})}}\right ),
\label{time23}
\end{eqnarray}
which behaves like
\begin{eqnarray}
P(v,t)\sim e^{\gamma t}\Phi\left (\frac{v-v_{0}e^{-\gamma t}}{\sqrt{2D/\gamma}}\right ),
\label{time24}
\end{eqnarray}
for $t\rightarrow +\infty$. On the other hand, Eq. (\ref{time20}) takes the form
\begin{eqnarray}
P(v,t)=\sqrt{\frac{\gamma}{2\pi D (1-e^{-2\gamma t})}}e^{-\frac{\gamma (v-v_{0}e^{-\gamma t})^{2}}{2D(1-e^{-2\gamma t})}},
\label{time25}
\end{eqnarray}
which tends to the Maxwellian for $t\rightarrow +\infty$.

\subsection{The case $h(t)=t$}
\label{sec_ht}

\begin{figure}
\centering
\includegraphics[width=8cm]{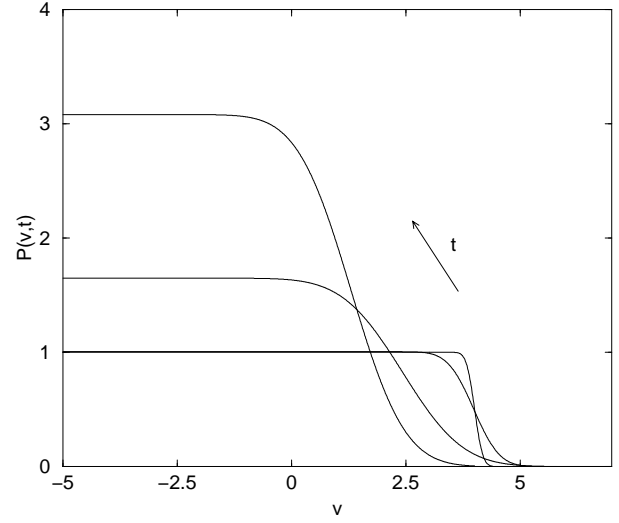}
\caption{The function $P(v,t)$ with $\gamma=D=1$ and $v_{0}=4$ for $t=0.01, 0.1, 1, 1.5$ starting from a step function $\eta(v-v_{0})$. }
\label{creneau}
\end{figure}

\begin{figure}
\centering
\includegraphics[width=8cm]{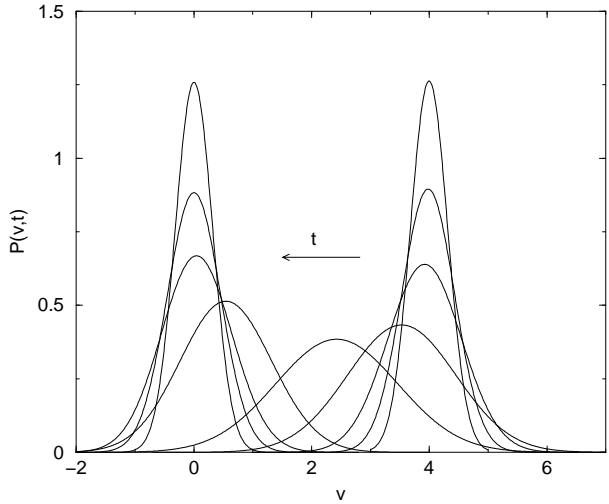}
\caption{The function $P(v,t)$ with $\gamma=D=1$ and $v_{0}=4$ for $t=0.05, 0.1, 0.2, 0.5, 1, 2, 3, 5, 10$ starting from $\delta(v-v_{0})$. }
\label{prob}
\end{figure}

In the case where the friction coefficient increases linearly with time ($h(t)=t)$, we get
\begin{eqnarray}
H(t)=\frac{t^{2}}{2}, \quad H_{1}(t)=e^{\gamma \frac{t^{2}}{2}},  
\label{time26}
\end{eqnarray}
\begin{eqnarray}
H_{2}(t)={D}\int_{0}^{t}e^{\gamma \tau^{2}}d\tau, \quad \chi^{2}(t)=De^{-\gamma t^{2}}\int_{0}^{t} e^{\gamma \tau^{2}}d\tau. \nonumber\\
\label{time27}
\end{eqnarray}
The last function can be written
$\chi^{2}(t)=\frac{D}{\sqrt{\gamma}}\phi(\sqrt{\gamma}t)$ where
\begin{eqnarray}
\phi(t)=e^{-t^{2}}\int_{0}^{t}e^{x^{2}}dx,
\label{dawson}
\end{eqnarray}
is the Dawson integral. It behaves like $\phi(t)\sim 1/(2t)$ for $t\rightarrow +\infty$ and like $\phi(t)\sim t$ for $t\rightarrow 0$.
Equation (\ref{time18}) then takes the form
\begin{eqnarray}
P(v,t)=e^{\gamma t^{2}/2}\Phi\left (\frac{v-v_{0}e^{-\gamma t^{2}/2}}{\sqrt{\frac{4D}{\gamma^{1/2}}\phi(\sqrt{\gamma}t)}}\right ),
\label{time28}
\end{eqnarray}
which behaves like
\begin{eqnarray}
P(v,t)\sim e^{\gamma t^{2}/2}\Phi\left (\frac{\gamma t (v-v_{0}e^{-\gamma t^{2}/2})}{\sqrt{2D}}\right ),
\label{time29}
\end{eqnarray}
for $t\rightarrow +\infty$. The time evolution of the profile $P(v,t)$
is represented in Fig. \ref{creneau}.  On the other hand,
Eq. (\ref{time20}) takes the form
\begin{eqnarray}
P(v,t)=\sqrt{\frac{\gamma^{1/2}}{4\pi D \phi(\sqrt{\gamma} t)}}e^{-\frac{\gamma^{1/2} (v-v_{0}e^{-\gamma t^{2}/2})^{2}}{4D\phi(\sqrt{\gamma}t)}}.
\label{time30}
\end{eqnarray}
For $t\rightarrow +\infty$, it behaves like
\begin{eqnarray}
P(v,t)=\left (\frac{\gamma t}{2\pi D}\right )^{1/2} e^{-\frac{\gamma t}{2D} (v-v_{0}e^{-\gamma t^{2}/2})^{2}},
\label{time31}
\end{eqnarray}
and eventually tends to $\delta(v)$. The system collapses to a Dirac
in velocity space due to the divergence of the friction coefficient
for $t\rightarrow +\infty$.  Alternatively, for $t\rightarrow 0$, the
initial distribution $\delta(v-v_{0})$ deforms itself according to
\begin{eqnarray}
P(v,t)=\frac{1}{\sqrt{4\pi D t}} e^{-\frac{(v-v_{0}e^{-\gamma t^{2}/2})^{2}}{4Dt}}.
\label{time32}
\end{eqnarray}
 The time evolution of the profile $P(v,t)$
is represented in Fig. \ref{prob}.

\subsection{Another example}
\label{sec_ext}

For completeness, we also provide the analytical solution of 
the equation 
\begin{eqnarray}
\label{ext1}
\frac{\partial P}{\partial t}=\frac{\partial}{\partial v}\left (D\frac{\partial P}{\partial v}+\gamma h(t) P\right ),
\end{eqnarray} 
where $D$ and $\gamma$ are constant parameters. Writing this equation in the form
\begin{eqnarray}
\label{ext2}
\frac{\partial P}{\partial t}-\gamma h(t)\frac{\partial P}{\partial v}=D\frac{\partial^{2}P}{\partial v^{2}},
\end{eqnarray}
we see that the second term is similar to an advection (in velocity space) by an effective velocity field $V(v,t)=-\gamma h(t)$. Let $v_{f}(t)$ be a solution of $dv_{f}/dt=-\gamma h(t)$. We take $v_{f}(t)=-\gamma H(t)$ where $H(t)$ denotes a primitive of $h(t)$ with $H(0)=0$. Then, we define $z=v-v_{f}(t)$ and $P(v,t)=\phi(z,t)$. Substituting these relations in Eq. (\ref{ext2}), we find that $\phi(z,t)$ satisfies the diffusion equation
\begin{eqnarray}
\label{ext3}
\frac{\partial \phi}{\partial t}=D\frac{\partial^{2}\phi}{\partial z^{2}}.
\end{eqnarray}
Using for example the results of Sec. \ref{sec_gen}, the general solution of this equation is
\begin{eqnarray}
\label{ext4}
\phi(z,t)=\frac{1}{\sqrt{\pi}}\int_{-\infty}^{+\infty}e^{-x^{2}}\phi_{0}(z-2\sqrt{Dt} x)dx.
\end{eqnarray}
Returning to original variables, we get
\begin{eqnarray}
\label{ext5}
P(v,t)=\frac{1}{\sqrt{\pi}}\int_{-\infty}^{+\infty}e^{-x^{2}} P_{0}(v+\gamma H(t)-2\sqrt{Dt} x)dx,\nonumber\\
\end{eqnarray}
where $P_{0}(v)=P(v,0)$ is the initial value of the probability distribution.  
If $P_{0}(v)=\eta(v-v_{0})$ is a step function, we find that
\begin{eqnarray}
{P}(v,t)=\Phi\biggl
({v-v_{0}+\gamma H(t)\over 2\sqrt{Dt}}\biggr ). \label{ext6}
\end{eqnarray}
Alternatively, if $P_{0}(v)=\delta(v-v_{0})$, we obtain
\begin{eqnarray}
P(v,t)=\frac{1}{\sqrt{4\pi Dt}}e^{-\frac{(v-v_{0}+\gamma H(t))^{2}}{4Dt}}.
\label{ext7}
\end{eqnarray}

\section{Conclusion}
\label{sec_conclusion}

We have presented first elements of kinetic theory appropriate to
inhomogeneous systems with long-range interactions. Explicit results
have been obtained for the HMF model in the low temperature limit
$T\rightarrow 0$ where the particles perform harmonic motion around
$\theta=0$ with the same frequency.  These results show that the
description of {\it strongly} inhomogeneous systems is very different
from what we are used to in ordinary kinetic theory. In particular,
the Kubo formula is not valid in its usual form because the
auto-correlation function of the force does not decrease sufficiently
rapidly for large times. On the contrary, it has a striking
oscillatory behaviour that averages to zero over a period. On the
other hand, the Einstein relation is broken. Indeed, the friction
coefficient depends on time and diverges linearly as $t\rightarrow
+\infty$ due to the cumulative effects of friction as the particle
undergoes several oscillations.

These results are strikingly different from those obtained in the
homogeneous phase $T>T_{c}$ of the HMF model where the particles
follow linear trajectories with constant velocity in a first
approximation. In that case, the Kubo formula and the Einstein
relation are recovered \cite{vatt}. It would be interesting to
consider the case of the weakly inhomogeneous HMF model. This could be
studied perturbatively by expanding the relations
(\ref{in6})-(\ref{in7}) close to the critical point $T\rightarrow
T_{c}^{-}$ where $B\rightarrow 0$. Similar expansions have been
previously considered in \cite{vatt} in other circumstances. It will
be necessary to distinguish between open trajectories that are similar
to the situation for $T>T_{c}$ from closed trajectories that are
similar to the situation for $T\rightarrow 0$. It would also be
interesting to investigate how collective effects can modify these
results (as in the case $T>T_{c}$). In particular, close to the
critical point the correlation time diverges which complicates the
kinetic theory. 

We stress that our approach does not provide a full kinetic theory of
the inhomogeneous HMF model. Our main motivation was to mention
the difficulties with such a kinetic approach due to Markovian effects
and spatial inhomogeneities and to show the differences with more
familiar kinetic theories. Even if we do not obtain
many explicit predictions, an interesting aspect of our study is to
show that the usual kinetic relations (Kubo formula, Einstein
relation,...) can break down for strongly inhomogenous systems with
long-range interactions such as the HMF model.  Finally, we would 
like to mention that it is important to develop a kinetic theory for
the {\it inhomogeneous} HMF model because there are situations of physical interest
where the system is spatially inhomogeneous. On the other hand,  the bath can have a
complicated phase-space structure with a hierarchical cluster size
distribution as exemplified in the numerical simulations of Rapisarda
\& Pluchino \cite{rapisarda}. The kinetic theory should take into account these effects.

\end{document}